\documentclass{emulateapj}

\usepackage{amsmath}
\usepackage{xspace} 
\usepackage{graphicx}
\usepackage{txfonts}
\usepackage{natbib}

\def\lax {\ifmmode{_<\atop^{\sim}}\else{${_<\atop^{\sim}}$}\fi}  
\def\gax {\ifmmode{_>\atop^{\sim}}\else{${_>\atop^{\sim}}$}\fi}  
\def\gtorder{\mathrel{\raise.3ex\hbox{$>$}\mkern-14mu
             \lower0.6ex\hbox{$\sim$}}}

\def\IGR{IGR~J00291+5934}
\def\SAX{SAX~J1808.4-3658}
\def\J1751{XTE~J1751-305}
\def\be{\begin{equation}}
\def\ee{\end{equation}}

\usepackage{lscape}

\begin{document}

\title{Energy dependent $\sim 100~\mu s$  time lags as  observational
  evidence of Comptonization effects  in the neutron star plasma environment}

\author{Maurizio Falanga\altaffilmark{1,2} \& Lev Titarchuk\altaffilmark{3,4,5} }

\altaffiltext{1}{ CEA Saclay, DSM/DAPNIA/Service d'Astrophysique (CNRS FRE 
  2591), F-91191, Gif sur Yvette, France; e-mail: mfalanga@cea.fr}
\altaffiltext{2}{Unit\'e mixte de recherche Astroparticule et
  Cosmologie, 11 place Berthelot, 75005 Paris, France} 
\altaffiltext{3}{George Mason University/Center for Earth Observing
  and SPace Research, Fairfax, VA 22030; and US Naval Research
  Laboratory, Code 7655, Washington, DC 20375-5352; e-mail:
ltitarchuk@ssd5.nrl.navy.mil}
\altaffiltext{4}{Dipartimento di Fisica, Universita di Ferrara, Via Saragat 1, I-44100 Ferrara, Italy e-mail:
titarchuk@fe.infn.it}
\altaffiltext{5}{Goddard Space Flight Center, NASA, Exploration of the
Universe Division, code 661, Greenbelt MD 20771; e-mail:
lev@milkyway.gsfc.nasa.gov}

\begin{abstract}

We present a  Comptonization model for the observed  properties of the
energy-dependent soft/hard  time lags and  pulsed fraction (amplitude)
associated with the pulsed emission of a neutron star (NS). We account
for  the soft lags  by downscattering  of hard  X-ray photons  in the
relatively cold  plasma of the disk  or NS surface. A  fraction of the
soft X-ray photons  coming from the disk or  NS surface upscatter off
hot electrons in the accretion  column. This effect leads to hard lags
as a result of thermal  Comptonization of the soft photons. This model
reproduces  the observed  soft  and hard  lags  due to  the down-  and
upscattered radiation as a  function of the electron number densities
of the  reflector, $n_{e}^{\rm ref}$,  and the accretion  column, $n_{
e}^{\rm  hot}$.  In  the  case of  the  accretion-powered  millisecond
pulsars \IGR,  \J1751, and  \SAX\, the observed  time lags  agree well
with  the  model.  Soft   lags  are  observed  only  if  $n_e^{ref}\ll
n_e^{hot}$. Scattering of  the pulsed emission  in the NS environment  
may account for the observed time lags as a non-monotonic function of 
energy.  The time lag measurements can be used as a probe of the innermost 
parts of the NS and accretion disk. We determine the upper and lower limits of
the density  variation in  this region using  the observed  time lags.
The observed  energy-dependent pulsed amplitude  allows us to  infer a
variation of the  Thomson optical depth of the  Compton cloud in which
the accretion column is embedded.

\end{abstract}


\section{Introduction}

Accreting neutron star and millisecond pulsar (MSP) source
spectra are successfully fitted by a two-component model 
consisting of Comptonization of higher temperature blackbody photons
and thermal soft X-ray emission. The emergent radiation is presumably
produced by thermal and dynamical  Comptonization of seed photons
coming  from the NS surface and an accretion disk [for NS:
  e.g. \citep []{tmk96,tmk97,TS05,paiz} 
and MSP:  e.g. \citep[] {gilfanov98,gdb02,gp05,mfa05,mfb05}]. 
 
Additional information for the X-ray production processes and
emission environment can be obtained by studying the pulse profile
and phase shift between X-ray pulses at different energy ranges. For
\SAX\ and later for \J1751\ it was found that the low-energy pulses
lag behind the high-energy pulses (soft phase/time lags) monotonically
increasing with energy  and saturating at about 10--20 keV
\citep{CMT98,Ford00,gp05}. 

This time lag effect was first interpreted  as a result of 
photon delay due to downscattering of  hard X-ray photons in the
relatively cold plasma of the disk or NS surface \citep{CMT98,TCW}.
It was argued that the photon time lags were an intrinsic signature of the
interaction of the Comptonized radiation with the NS and accretion
disk plasma. Moreover, the absolute values of the time lags (about
hundreds $\mu$s) are consistent with the electron scattering time
scale $t_{\rm C}=\tau_{\rm T}(L/c)$.  Effective Thomson  optical depth
of the cold reflector $\tau_{\rm T}$
is of the order of a few. Typical sizes of the NS photosphere and half-width
of the disk $L$ are of the  order of $10^{6}$ cm. On the other hand,
\citet{pg03} suggested that the lags may be produced by a combination
of different angular distributions of the radiation components  and
relativistic effects.  Recently,  soft lags were also found up to
$\sim100$ keV in the 1.67 ms accreting MSP \IGR, where a more complex
energy dependence was revealed  \citep{mfa05}.

In this Letter, we present a model for  the energy-dependent time
lags as a result of photon downscattering and upscattering in a
Comptonized medium.  We show that the resulting time lag signs
and  behavior as a function of energy  depend on the ratio of
densities of the Compton emission area, NS surface area and accretion
disk regions. By time lag, we mean the time or phase shifts
  between X-ray pulses at different energies.

\begin{figure}
\centering
\includegraphics[width=5.0cm,angle=90]{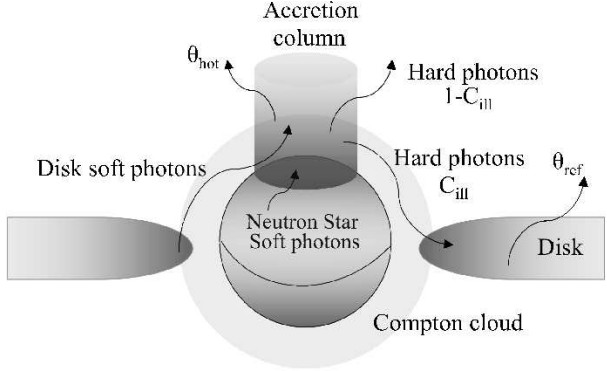}
\caption{This cartoon illustrates the different emission patterns
  responsible for the time lags of the pulsed emission where
  $\theta_{\rm e}^{\rm hot}$ and $\theta_{\rm e}^{\rm ref}$  are the
  dimensionless temperatures of the accretion column and reflector,
  respectively. $C_{\rm ill}\sim 0.1$ is the disk illumination
  fraction.  Soft time lags of the pulsed emission are the result of
  downscattering of hard X-ray photons in the relativly cold plasma of
  the disk.  The fraction of hard X-ray photons ($1-C_{\rm ill}$)
  directly seen by the observer is related to the  upscattered soft
  photons coming from the NS and disk. We suggest that  the emergent
  pulsed flux (pulsed fraction) is formed  in the Compton cloud.  
}
\label{fig:geom}
\end{figure}

\section{Comptonization model of energy-dependent time lags}
\label{sec:Model}

In Figure  \ref{fig:geom}, we present a typical scenario of the
formation of the X-ray spectrum of a NS binary. We assume that  the
hard radiation is produced in  an  accretion column or Compton cloud
(Comptonization  region) and illuminates the     relatively cold
material of the NS and  disk. The emergent hard  X-ray emission
is  a result of the upscattering
of NS and  disk soft photons off of hot electrons in the
Comptonization region and the reflection of  this upscattered radiation
from the NS and disk surfaces.  
The physical parameters of the source, the geometrical and
  optical thickness 
of the disk, the disk plasma conductivity and the NS magnetic field determine
the position of the inner edge of the disk and the illumination covering
factor $C_{\rm ill}$. 
The fraction of the hard photons
intercepted by the disk and NS and ultimately reflected ($C_{\rm ill}$)
can be in the range of at most 10--15\%,  depending on the details of the
geometry.   The parameter $C_{\rm ill}$  is a  product of the albedo,
$A\sim 0.3$, of the cold material \citep{basko74}  and  the fraction of the
Comptonized photons,  $f_{\rm int}$, emitted by the isotropically radiated
central source  and then intercepted by the infinite disk. The
fraction $f_{\rm int}$ is less then 25\% \citep{LST85}.  Thus
$C_{\rm ill}=A\cdot f_{\rm int}\lax0.1$. In Sec. \ref{sec:fit_result}, we use 
$C_{\rm ill}$ = 0.1 as a representative value for low magnetized
($\sim 10^8$ G)
neutron stars where the innermost edge of the disk is just a few kilometers
from the NS surface.  

The photon energy change 
 per electron scattering $\delta E/\delta u$ depends on the photon
energy, $E$, and  electron plasma temperature $kT_e$:
\begin{equation}
\frac{1}{z}\frac{\delta z}{\delta u}=4\theta-z,
\label{en_perscat}
\end{equation}
where $z=E/m_ec^2$ is the dimensionless photon energy  and 
$\theta=kT_e/m_ec^2$ is the dimensionless plasma temperature with
respect to the electron rest energy  $m_ec^2$ \citep{ST80}. 
In a cold plasma (NS photosphere or accretion disk) for which the plasma
temperature  $kT_e$ is much less than a given energy of the photon $E$,
the solution of the differential equation (\ref{en_perscat}) is 
\begin{equation}
u=\frac{1}{4\theta}\ln\frac{1-4\theta/z_{\ast}}{1-4\theta/z}
\label{sol_eq}
\end{equation}
where $z_{\ast}$ is the dimensionless energy of a given photon after
$u$  scatterings for which the initial dimensionless energy is $z$,
$u=(n_e\sigma_{\rm T}c)t$  is the dimensionless time measured in units of mean time
between scatterings (number of scatterings),  $n_e$ is the plasma
number density, and $\sigma_{\rm T}$ is the Thomson electron
cross-section. In a hot plasma, e.g. the Comptonization region,  
$kT_{\rm e}$, is much higher than the seed photon energy
$E_{\rm s}$.  In this case, the solution of Eq. (\ref{en_perscat}) is 
\begin{equation}
u=\ln (z/z_s)/4\theta.
\label{hot_plasma}
\end{equation}
The appropriate solution of the exact kinetic equation \citep{TL95} 
gives  
\begin{equation}
u=\frac{\ln z/z_s}{\ln[1+(3+\alpha)\theta]},
\label{hot_plasma_keq}
\end{equation}
where $\alpha$ is the energy spectral index.   One can approximate Eq. 
(\ref{hot_plasma_keq}) by Eq. (\ref{hot_plasma}) if it is assumed that
$(3+\alpha)\;\theta\ll1$ and $\alpha\lax 1$. Note that
Eq. (\ref{en_perscat}) is obtained using a diffusion approximation.  

The resulting time lags are  a linear combination of the
positive (hard) time lags formed in the Comptonization emission area and
the negative (soft) ones formed in the accretion disk and NS photosphere as a
result of reflection of the hard radiation (see Fig. \ref{fig:geom}): 
$$
\Delta t  =  -\frac{C_{\rm ill}}{\sigma_{\rm T}n_{\rm e}^{\rm ref} c} \times
$$
\begin{equation}
\biggl[\frac{1}{4\theta_{\rm ref}}\ln
\frac{1-4\theta_{\rm
      ref}/z}{1-4\theta_{\rm ref}/z_{*}}
      -\frac{n_{\rm
  e}^{\rm ref}}{n_{\rm e}^{\rm hot}}\frac{1-C_{\rm ill}}{C_{\rm ill}}
\frac{\ln(z/z_{*})}{\ln[1+(3+\alpha)\theta_{\rm hot}]}\Biggr]
\label{res_timelag}
\end{equation}
where $n_{\rm e}^{\rm ref}$ is the electron number density of the
reflector, $n_{ e}^{\rm hot}$ is the electron number density of the
Comptonization emission area  (accretion column) and $\theta_{\rm
ref}=kT_e^{\rm ref}/(m_{e} c^2)$ is the dimensionless temperature of the
reflector.   We assume a typical value of $\theta_{\rm ref} <
0.7$~keV/511~keV.    $kT_{\rm e}^{\rm hot}$ and $\alpha$  are the
best-fit 
parameters for the hot plasma temperature and spectral index of the
Comptonization spectrum respectively, and $\theta_{\rm  hot}=kT_{
e}^{\rm hot}/(m_{ e} c^2)$.

We also assume that the seed photon energy is near
the lowest energy of the downscattered photons, i.e. $z_s\approx z_{\ast}$
(see Eqs. \ref{sol_eq}, \ref{hot_plasma_keq}).   
The observational value of $E_{\ast}$  is  about  3 keV. We
set the  value of  the illumination factor $C_{\rm ill}$ at
about 0.1 (see above for details). 

The essential model  parameters for fitting the observed  lags are
the number  densities $n_{e}^{\rm ref}$ and  $n_{ e}^{\rm
hot}$. Eq. (\ref{res_timelag}) is insensitive to $\theta_{\rm
ref}$ if $\theta_{\rm ref}<z_{\ast}/4$. The first term in parentheses
of Eq. (\ref{res_timelag}) is related to the downscattering lags,
and the second term is related to the upscattering lags. The
positive (hard) lags cannot be observed if the density of the hot
plasma is much 
higher than that of the ``cold''  reflector.  This is because  the 
mean free path between consecutive scatterings $l=1/(n_e\sigma_{\rm
T})$ and therefore the time lag $\Delta t=u l/c$ are  inversely
proportional to the number density $n_e$. Consequently, the photons
spend less time $\Delta t$ in the higher density plasma than in the
lower density plasma to undergo the same number of scatterings.

The observer detects the negative (soft) lags only if
$n_e^{ref}\ll n_e^{hot}$ and thus the relative fraction of the
negative (downscattering)  time lags in the apparent time lag sum is 100\%
(i.e. $C_{\rm ill}=1$). 
In this case, the observed absolute values of
the soft time lags  monotonically increase with  energy.
Using all of the conditions described above Eq. \ref{res_timelag}
  can be written, after some algebra, as 
\begin{equation}
\Delta t  \approx - \frac{1}{\sigma_{\rm T}n_{\rm e}^{\rm ref} c}
\biggl(\frac{1}{z_{\ast}}-\frac{1}{z}\biggr).
\end{equation}
This type of time lag correlation with energy  was found for the
first time in the accreting MSP \SAX\ \citep[see][]{CMT98}. 

\begin{figure}
\centering
\includegraphics[width=8.cm]{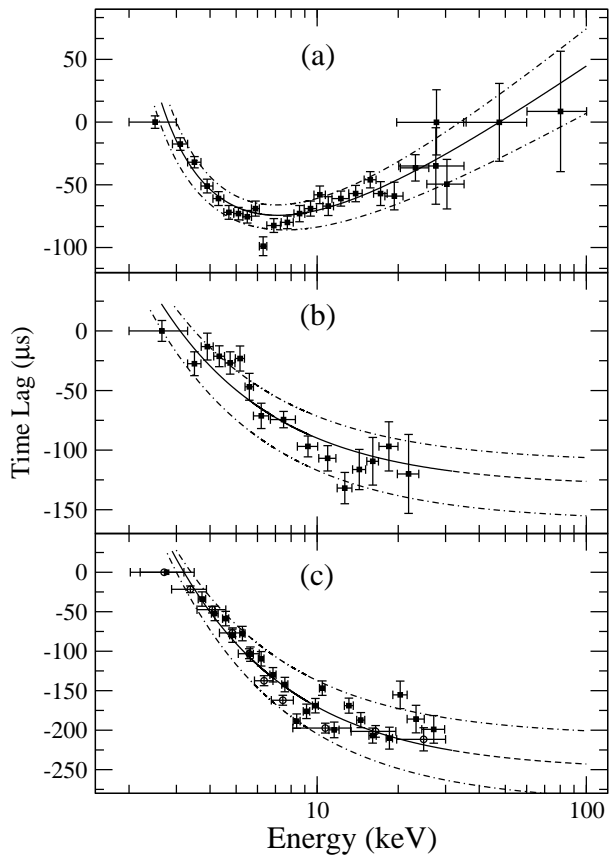}
\caption{The measured soft time lags  of the pulse profile versus energy
 (crosses) with respect to the first energy channel. The best-fit
 curve using the Comptonization model (see Sec. \ref{sec:Model}) is
 shown with the solid line.   The dot-dashed lines in panel (a)
 correspond to the upper and lower limits of the electron number
 densities of the Comptonization emission area, $n_{\rm e}^{\rm hot}$
 and  of  the reflector, $n_{\rm e}^{\rm ref}$ in \IGR.  The panels
 (b) and (c)  show \J1751\ and \SAX, respectively.  The
 dot-dashed lines correspond to the upper and lower limits of  $n_{\rm
 e}^{\rm ref}$.  }
\label{fig:bestfit}
\end{figure} 

\section{Application  of the model  to observed time lags}
\label{sec:fit_result}

To apply the Comptonization model
Eq. (\ref{res_timelag}) to observed time lags, we used the observed
pulse phase lag data and 
best-fit spectral parameters  $kT_{\rm e}^{\rm hot}$  and $\alpha$ of
the accreting MSPs \IGR, \J1751, and \SAX. 
Figure \ref{fig:bestfit} shows, from top to bottom, \IGR, \J1751, and
\SAX\ \citep{mfa05,gp05, CMT98,Ford00}. The
  measurements were made either through 
computing cross-power spectra between different energy bands or
through cross-correlating the folded pulse profile in different energy
bands.

Figure  \ref{fig:bestfit} shows
the best-fit curves using the Comptonization model described in
Sec. \ref{sec:Model}. We used $kT_{\rm e}^{\rm ref} = 0.4$ keV as the
 reflector temperature for all  fits.  The  model for fitting the
time lags in \J1751\ and \SAX\  has only one free parameter, the
number density of the ``cold'' reflector $n_{\rm e}^{\rm ref}$ (see
Sec. \ref{sec:Model}). Presumably, in \J1751\ and \SAX\ the density of
the Comptonization region is much higher than that of the ``cold''
reflector, i.e. $n_{\rm e}^{\rm hot}\gg n_{\rm e}^{\rm ref}$. This may
indicate that the seed photons are  Comptonized
(upscattered) in the very dense plasma of the accretion column
\citep[see e.g.][]{BS76,BW06}. The best-fit values are $n_{\rm e}^{\rm
ref}= 6.3\times10^{19}$ cm$^{-3}$ and $3.2\times10^{19}$ cm$^{-3}$
for \J1751\ and \SAX, respectively. The fits of the time lag data for
\IGR\ provide us with best-fit values for the ``cold'' and hot plasma
densities. Both the positive (upscattering) and negative
(downscattering) time lags  contribute to the apparent time lag
because $n_{\rm e}^{\rm ref}= 6.9\times10^{18}$ cm$^{-3}$ and $n_{\rm
e}^{\rm hot}=  2.1\times10^{18}$ cm$^{-3}$ are of the same order of
magnitude.

The time lag  data of  \IGR, \J1751, and \SAX\ were collected in
different time intervals lasting from hours to days
\citep{CMT98,Ford00,gp05,mfa05}. However, a hydrodynamical  (density
perturbation) time,  $t_{hydro}$, in the innermost part of the X-ray
NS source is of the order of the ratio of the NS radius to  the sound
speed, that is,  $t_{hydro}\sim R_{\rm NS}/c_{\rm sound}\gax 0.1$ s.
Thus, during the data collection periods, the densities of the
surrounding plasma can vary. As a result, the time lags also vary
since they are sensitive to density variations (see
Eq. \ref{res_timelag}). Therefore, we can infer  these density
variations from the time lag data. We construct a corridor 
between two curves of time lag vs energy which includes all of the time
lag data points 
(see Fig. \ref{fig:bestfit}). This allows us to constrain
$n_{\rm e}^{\rm ref}$ and $n_{\rm e}^{\rm hot}$.   The lower and upper
curves in Figure \ref{fig:bestfit} correspond in panel (a) to $n_{\rm e}^{\rm
ref}=  (6.1-8.0)\times10^{18}$ cm$^{-3}$ and  $n_{\rm e}^{\rm hot} =
(1.6-2.6)\times10^{18}$ cm$^{-3}$, respectively.  For the sources in panels (b)
\J1751\ and (c) \SAX\, the density variations are  $n_{\rm e}^{\rm
ref}= (6.0-6.6)\times10^{19}$ cm$^{-3}$ and  $n_{\rm e}^{\rm ref}=
(2.9-3.6)\times10^{19}$ cm$^{-3}$.    Thus, the plasma density of the
``cold'' reflector can change as much as 10 \% during the entire data
collection.

\begin{figure}
\centering
\includegraphics[width=8.cm]{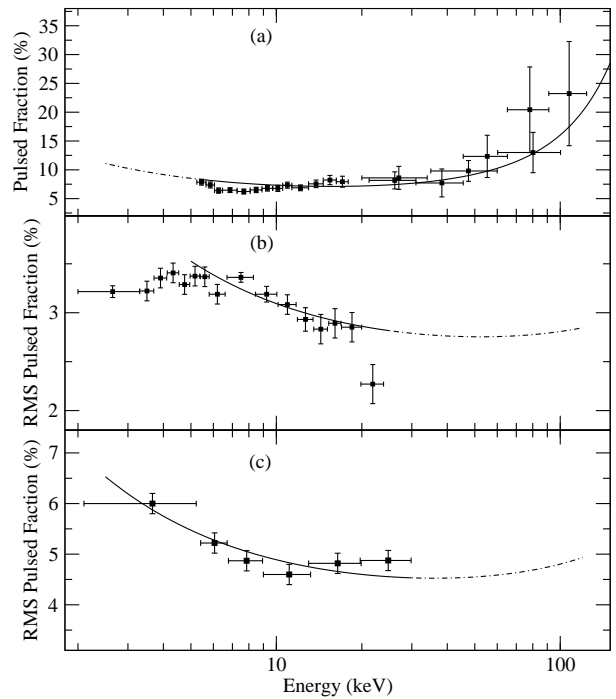}
\caption{The observed  energy dependent pulsed fraction of (a) \IGR\ ,
(b) XTE~J1751-305, and (c) SAX~J1808.4-3658 along with the best-fit
model (see Sec. 4).
}
\label{fig:pf}
\end{figure}

\section{Pulsed fraction of X-rays: observation and theory}
\label{sec:fit_fract_result}
In Figure \ref{fig:pf}, we show the energy dependent pulsed fraction of
(a) \IGR, (b) \J1751\ and (c) \SAX. The pulsed fraction of \IGR\ vs
energy is explained by the energy  dependent  electron
cross-section $\sigma_e(E) =\sigma_{\rm T}(1-2z)$ \citep[see
e.g.][]{pom} and consequently by Compton cloud  optical depth as a
function of energy
\begin{equation}
\tau_{\rm cl}(E)=\tau_{{\rm T, cl}}(1-2z).
\label{electron_tau}
\end{equation}
We assume that the accretion column is embedded in the Compton
cloud  as shown in Figure \ref{fig:geom}  \citep[see also][for the
geometry details]{TCW}.
Because $\tau_{\rm cl}(E)$ decreases with energy,  a larger fraction
of the pulsed direct hard X-rays which originated  in the accretion
column can escape to the observer. This fraction is given by
\begin{equation}
A_{\rm rms,es}(E)=A_{\rm rms,es}(0)\exp[-\tau_{\rm cl}(E)]
\label{A-fraction}
\end{equation}

A different scenario of  energy dependent amplitude  formation
can be if  there is no electron (Compton) cloud
between the accretion column (where the Comptonization spectrum is
formed) and the observer. In this case  the  energy dependence of the amplitude
is formed as a result of   upscattering.    The soft photons
upscattered  off hot electrons (in the accretion column)   increase their
energy  with the number of scatterings.  On the other hand, the amplitude
of the pulsed radiation exponentially decreases  with the number of
scatterings $A_{\rm rms}(u)=A_{\rm rms}(0)\exp(-\beta u)$, where
$\beta$ is the inverse of the average number of scatterings \citep[see
e.g.][]{ST80} and  consequently with an energy of the upscattered
photon (see Eq. \ref{hot_plasma})
\begin{equation}
A_{\rm rms,up}[u(E)]=A_{\rm rms,up}(E_s)(E/E_s)^{-\alpha_{\rm cl}}
\label{A-upscatter_fraction}
\end{equation}
 (where $\alpha_{\rm cl}=\beta/4\theta$ is approximately the energy  index
 of the Comptonization spectrum).  A characteristic seed (disk and NS)
 photon energy  is $E_s< 3$ keV.  Consequently,
 all photons at  higher  energies ($>3$ keV) are produced by
 upscattering, and their amplitude should decrease  with energy.

In the general case,  some  of the pulsed  soft photons, (a fraction
$A_{\rm rms,up}$),  may upscatter  off hot electrons in the accretion
column on the way out as other pulsed photons (a fraction
$A_{\rm rms,es}$) forming the hard X-ray tail may escape to the
observer while passing through the Compton cloud.
In this  case, the following combination of Eqs. (\ref{A-fraction} and
\ref{A-upscatter_fraction})
\begin{equation}
A_{\rm rms}(E)= A_{\rm rms,up}(E/E_s)^{-\alpha_{\rm cl}} + A_{\rm
rms,es}e^{-\tau_{\rm cl}(E)}
\label{A-general_fraction}
\end{equation}
lead to the formula of the emergent pulsed amplitude.

In Figure \ref{fig:pf}  we present  the best-fit model (see
Eq. \ref{A-general_fraction})  along with the data.  We  fit only
those data points which correspond to  energies higher than the seed
photon energy.   We find for (a) \IGR\
$\tau_{\rm T, cl} = 3\pm0.2$ and $A_{\rm rms,es} = 98^{+2}_{-6}$ \%,
  $A_{\rm rms,up} = 5.2\pm0.9$ \%. For (b) $\tau_{\rm T, cl} =
  0.25\pm0.09$ and $A_{\rm  rms,es} = 3.5\pm0.8$ \%,  $A_{\rm rms,up}
  = 1.8\pm1.1$ \% and (c) $\tau_{\rm T, cl} = 0.4\pm0.2$ and $A_{\rm
  rms,es} = 5.9\pm6$ \%, $A_{\rm rms,up} = 2.5\pm1.2$ \%,
  respectively. For all of the fits $\alpha_{\rm cl}$ was $\sim0.8$.

\section{Discussion and Conclusions}
\label{sec:DisCon}

In this Letter we have shown that the short time lags, phase shifts of
the  order of 100 $\mu$s  
between X-ray pulses at different energy ranges,  detected in the
accreting MSPs are the result of photon 
upscattering and downscattering in the plasma environment of the NS.
In our model the Comptonization emission area, NS surface, and
accretion disk are responsible for the time lag formation.  The photon
time lag correlations as a function of the electron 
densities of the reflector, n$_{\rm e}^{\rm ref}$, and the accretion
column, n$_{\rm e}^{\rm hot}$,  at a given energy are successfully
reproduced for a reasonable choice of  model ($C_{\rm ill}, kT_{\rm
e}^{ref}$) and best-fit photon spectral parameters ($\alpha, kT_{\rm
  e}^{hot}$). 
As illustrated in the schematic view of the accretion flow
(see Fig. \ref{fig:geom}), the hard X-ray photons are downscattered in
the relatively cold plasma of the disk  and therefore  produce  the
observed soft lags.
The observed hard time lags are  a result of thermal
upscattering of soft photons coming from the disk and NS off hot
electrons of the Comptonized region.  This conclusion is based on the spectral
shape of the emergent spectrum and time lag behavior as a function of
photon energy for three MSP sources. Indeed, we find for \IGR\ with a
$\Delta t_{max}\sim 100$ $\mu$s that $n_{\rm  e}^{\rm ref} \sim
3\times n_{\rm e}^{\rm hot}$, indicating that the densities of the
relatively ``cold'' disk plasma and of the
accretion column are on the 
same order of $10^{18}$ cm$^{-3}$ for  $C_{\rm ill}\sim 0.1$.  For \SAX\
and \J1751\ the observed energy-dependent time lags
need disk density of a factor of seven higher. The hard lags of the
emission due 
to Compton upscattering are negligible because of the expected 
compactness of the accretion column region.

However, one can notice in Figure \ref{fig:bestfit}  that there are
 signs of turnover (saturation) for time lags vs energy  at $\sim$20 keV
 for \SAX\ and \J1751 which are  similar  to  that in \IGR\ which has a
 turnover at about 7 keV. The high values of $n_{\rm e}^{\rm hot}$
 with respect to  $n_{\rm e}^{\rm ref}$  would not allow us to see the
 contribution of the hard lags in the resulting time lag vs energy plot.
 
 It is also important to emphasize that the energy-dependent time lags
 allow us to reveal the plasma density variation within 10\%  near the
 NS.  It implies that the mass accretion rate and
 ultimately the observed luminosity should also vary within 10\%.
 Thus,  we can conclude  that the observed variability of X-ray
 radiation in the compact objects is presumably driven by this
 variation of the density of the mass accretion there.

Measurements of time lags have been done for black hole binaries
\citep[e.g.][]{KK87,M95,N99}.
Making a comparison between these measurements and that of time/phase
shifts between X-ray pulsations in accreting MSPs is difficult. In the
case of black hole binaries, the time lag is generally calculated in two
different energy bands and is related to different Fourier frequencies.
For black holes, the Fourier frequency dependence can be due to
hydrodynamical disk instability, mass accretion configuration, or
quasi-periodic instability. This work with accreting
millisecond NSs provides a unique advantage since the X-ray pulsations
give a perfect clock which allows to determine precise information about
the photon delay at different energy bands.

\begin{acknowledgements} 
 We are grateful to Philippe Laurent for helpful discussions and Erin W.
 Bonning for careful reading of the manuscript.  MF
acknowledges the French Space Agency (CNES) and LT the University of
Ferrara for financial support of this work.  We are grateful to the referee
 for his/her interesting comments regarding the paper content. 
\end{acknowledgements}

\end{document}